\documentstyle[12pt]{article}
\begin{document}

\begin{center}
{\Large \bf Causal Theory for the Gauged Thirring Model}
\end{center}
\vspace{1.2cm}
\begin{center}
L. A. Manzoni and B. M. Pimentel\linebreak[1] \linebreak[1]
Instituto de F\'{\i}sica Te\'{o}rica\linebreak[1]
Universidade Estadual Paulista\linebreak[1]
Rua Pamplona, 145\linebreak[1]
01405-900 - S\~ao Paulo, S.P. \linebreak[1]
Brazil\linebreak[1]
\linebreak[1]
and
\linebreak[1]
\linebreak[1]
J. L. Tomazelli\linebreak[1] \linebreak[1]
Departamento de F\'{\i}sica e Qu\'{\i}mica - Faculdade de Engenharia\linebreak[1]
Universidade Estadual Paulista - Campus de Guaratinguet\'a\linebreak[1]
Av. Dr. Ariberto Pereira da Cunha, 333\linebreak[1]
12500-000 - Guaratinguet\'a, S.P. \linebreak[1]
Brazil
\end{center}

\vspace{2.2cm}
\begin{center}
{\bf Abstract}
\end{center}

We consider the ($2+1$)-dimensional massive Thirring model as a gauge theory, with one fermion flavor, in the framework of the causal perturbation theory and address the problem of dynamical mass generation for the gauge boson. In this context we get an unambiguous expression for the coefficient of the induced Chern-Simons term.
\vspace{0.5cm}


keywords: Thirring model; mass generation; causality

\pagebreak
\section{Introduction}

\hspace{0.7cm}Recently there has been renewed interest in theories involving four-fermion interactions as a framework to the study of the top quark condensate\cite{hill}. In these models, based in the Nambu and Jona-Lasinio model\cite{njl}, the top quark acquires mass when the four-fermion coupling constant is larger than a certain critical value. It has been shown that a similar behavior occurs in $d$ ($2\leq d<4$) dimensional Thirring-like four-fermion interactions\cite{gomes}-\cite{kon}.

The Thirring model\cite{th} was originally proposed as a soluble model for interaction of fermions in ($1+1$) dimensions. This original version has no local gauge symmetry. Since then, several authors\cite{gomes}-\cite{hands}, using a linearized version of the model by introducing an auxiliary vector field, have studied the Thirring model in $d$ dimensions, in the context of $1/N$ expansion, taking into account a ``gauge fixing'' term. However, as showed in \cite{kk}, \cite{ito} and \cite{kon}, one can implement gauge invariance by using the St\"{u}ckelberg formalism, so that the Thirring model emerges as a gauge-fixed version of a gauge theory. This model has been used to study fermion dynamical mass generation\cite{ito}\cite{kon}\cite{sug}.

In ($2+1$) dimensions these models exhibit a richer structure. Namely, for odd number of  massive 2-component fermions (the fermion masses may be present in original Lagrangian or have dynamical origin) a Chern-Simons parity-breaking term is induced. This term is relevant to the quantum Hall effect\cite{hall} and, in particular, to dynamical mass generation for the associated vector field. 

In this paper, we will consider the 3-dimensional Thirring model as a gauge theory for one fermion flavor. The approach we will adopt to study this model was proposed by Epstein and Glaser\cite{eg} in the 70's and later applied to QED by Scharf \cite{sch}. Their method, in which causality plays a central role, has the advantage that all physical quantities are mathematically well defined and ultraviolet divergencies do not occur if one carefully carries out the splitting of distributions in the perturbation series. In particular, this implies that no ultraviolet cut-off has to be introduced and, for a nonrenormalizable theory, this is the interesting point.

Our main interest here is the coefficient of the induced Chern-Simons term, because is generally stated that this coefficient is dependent on the regularization scheme used to treat the divergencies\cite{kon}\cite{djt}, even though this difficulty has already been overcomed in usual treatments of QED$_3$\cite{pst}\cite{pt}. As showed in \cite{swpt}, the causal approach affords an unambiguous method for dealing with such problems.

This paper is organized as follows. In section 2 we introduce the gauged version of the Thirring model\cite{ito}\cite{kon}, for one fermion flavor, using the St\"{u}ckel{\-}berg formalism. In section 3 we give a brief presentation of the method of Epstein and Glaser. A proof of the nonrenormalizability of the Thirring model in the context of the causal method is given in section 4. In section 5 we obtain the vacuum polarization tensor and the modified gauge boson propagator, showing that there is generation of dynamical mass for the gauge boson as a function of  the coupling constant. Section 6 is devoted to conclusions. 

\section{Thirring Model as a Gauge Theory}

Following closely refs. \cite{ito} and \cite{kon}, we present in this section the Thirring model as a gauge theory. In (2+1) dimensions the Lagrangian density for the massive Thirring model with one fermion flavor is

\begin{equation}
{\cal L} = \overline{\psi}i\gamma^{\mu}\partial_{\mu}\psi  - m\overline{\psi}\psi -\frac{G}{2}(\overline{\psi}\gamma^{\mu}\psi)(\overline{\psi}\gamma_{\mu}\psi),
\label{l}
\end{equation}

\noindent
where $\psi$ is a 2-component fermion field with mass $m$, supposed to be positive. The coupling constant $G$ have dimension of $(mass)^{-1}$ and will be redefined as $G=\frac{e^2}{M^2}$, with $e$ a dimensionless parameter\cite{kon2}.

The  algebra for the $\gamma$ matrices in (2+1) dimensions is realized using the Pauli matrices

\begin{equation}
\gamma^0=\sigma_3, \hspace{0.4cm} \gamma^1=i\sigma_1, \hspace{0.4cm} \gamma^2=i\sigma_2,
\label{matgam}
\end{equation}

\noindent
with

\begin{equation}
\{\gamma^{\mu},\gamma^{\nu}\}=2g^{\mu\nu}, \hspace{0.4cm}
\gamma^{\mu}\gamma^{\nu}=g^{\mu\nu}-i\varepsilon ^{\mu\nu\delta}\gamma_{\delta},
\label{gcomut}
\end{equation}

\noindent
where $g_{\mu\nu}= diag (1, -1, -1)$ and $\varepsilon ^{\mu\nu\delta}$ is the totally antisymmetric Levi-Civita tensor.

We can linearize the four fermion interaction by the introduction of an auxiliary vector field $\tilde{A}_\mu$, so that the Lagrangian is rewritten as

\begin{equation}
{\cal L}^{'} = \overline{\psi}i\gamma^{\mu}\tilde{D}_{\mu}\psi  - m\overline{\psi}\psi +\frac{M^2}{2}\tilde{A}^{\mu}\tilde{A}_{\mu},
\label{ll}
\end{equation}

\noindent
with $\tilde{D}_{\mu}=\partial_{\mu}-ie\tilde{A}_{\mu}$. It is important to note that, in spite of  the formal similarity, $\tilde{D}_{\mu}$ is not a covariant derivative, because the Lagrangian density (\ref{ll}) doesn't have local gauge symmetry.  The 3-vector $\tilde{A}_{\mu}\equiv -\frac{e}{M^2}\overline{\psi}\gamma_\mu\psi$ is just a suitable representation of the current.

However,\hfill one\hfill can\hfill introduce\hfill a \hfill local\hfill gauge \hfill symmetry\cite{kk}\hfill making \hfill use\hfill of\hfill the \\ St\"{u}ckelberg formalism. Namely, we decompose the vector field $\tilde{A}_{\mu}$ according to $\tilde{A}_{\mu}=A_{\mu}-\partial_{\mu}\theta$, where $A_{\mu}$ is a vector field and $\theta$ a neutral scalar field, whereas we perform the change  $\psi \rightarrow e^{-ie\theta}\psi$, $\overline{\psi} \rightarrow \overline{\psi}e^{ie\theta}$ (for a review of St\"{u}ckelberg's formalism see \cite{fuj}). Thus we get

\begin{equation}
{\cal L}^{'{'}} = \overline{\psi}i\gamma^{\mu}D_{\mu}\psi  - m\overline{\psi}\psi +\frac{M^2}{2}(A_{\mu}-\partial_{\mu}\theta)^2,
\label{ls}
\end{equation}

\noindent
with $D_{\mu}=\partial_{\mu}-ieA_{\mu}$. This Lagrangian is invariant  under the gauge transformation

\begin{eqnarray}
A_{\mu}&\rightarrow& A^{'}_{\mu}=A_{\mu}+\partial_{\mu}\phi ,\nonumber \\ 
\theta &\rightarrow& \theta^{'}=\theta+\phi ,\label{gtran} \\ 
\psi &\rightarrow& \psi^{'}=e^{ie\phi}\psi ,\nonumber \\
\overline{\psi} &\rightarrow& \overline{\psi}^{\; '}=\overline{\psi}e^{-ie\phi} \nonumber
\end{eqnarray}

\noindent
so that $A_{\mu}$ is really a gauge field and $D_{\mu}$ a covariant derivative. From (\ref{gtran}) we can see that in the unitary gauge $\theta^{'}=0$ one recovers the Lagrangian (\ref{l}), i.e., the original Thirring model is just a gauge-fixed version of  (\ref{ls}).

Since the Lagrangian (\ref{ls}) has local gauge symmetry, by adding a gauge-fixing and a Faddeev-Popov ghost term ${\cal L}_{GF+FP}$, we can obtain the complete BRST invariant Lagrangian\cite{kon}

\begin{equation}
{\cal L}_{Th,G}={\cal L}^{'{'}}+{\cal L}_{GF+FP},
\label{lbrs}
\end{equation}

\noindent
where ${\cal L}_{GF+FP}$ can be chosen in form

\begin{equation}
{\cal L}_{GF+FP}= -i{\bf  {\delta}_B}\left[\overline{c}\left( F[A,\theta] + \frac{\xi}{2}B \right)
\right],
\label{lfp}
\end{equation}

\noindent
so that (\ref{lbrs}) is invariant under the BRST transfomation

\begin{eqnarray}
{\bf  {\delta}_B}\psi(x)&=&iec(x)\psi(x),\nonumber \\
{\bf  {\delta}_B}\theta(x)&=&c(x), \nonumber \\
{\bf  {\delta}_B}A_{\mu}(x)&=&\partial_{\mu}c(x), \nonumber \\
{\bf  {\delta}_B}\overline{c}(x)&=&iB(x),\\
{\bf  {\delta}_B}c(x)&=&0,\nonumber \\
{\bf  {\delta}_B}B(x)&=&0.\nonumber 
\end{eqnarray}

\noindent
In the above expressions ${\bf {\delta}_B}$ represents the nilpotent BRST transformation, $c(x)$ and $\overline{c}(x)$ are the Faddeev-Popov ghosts and $B(x)$ is the Nakanishi-Lautrup auxiliary field. 

When the functional $F[A,\theta]$ is linear in both $A_{\mu}$ and $\theta$  the ghost fields decouple from the matter fields and, in special, choosing the $R_{\xi}$  gauge $F[A,\theta]=\partial_{\mu}A^{\mu}+\xi M^2\theta$,  the St\"{u}ckelberg field also decouples. So the Lagrangian (\ref{lbrs}), after integration over $B$, takes the form\cite{ito}\cite{kon}

\begin{equation}
{\cal L}_{Th,G}={\cal L}_{A,\psi}+{\cal L}_{\theta}+{\cal L}_{gh},
\end{equation}

\noindent
where

\begin{eqnarray}
{\cal L}_{A,\psi} &=& \overline{\psi}i\gamma^{\mu}D_{\mu}\psi  - m\overline{\psi}\psi +\frac{M^2}{2}A_{\mu}A^{\mu}-\frac{1}{2\xi}(\partial_{\mu}A^{\mu})^2 ,\label{lfim}\\ \nonumber \\
{\cal L}_{\theta} &=& \frac{1}{2} (\partial_{\mu}\theta)^2-\frac{\xi M^2}{2} \theta^2 ,\nonumber \\ \nonumber \\
{\cal L}_{gh} &=& i\left[ (\partial_{\mu}\overline{c})(\partial^{\mu}c)- \xi M^2 \overline{c}c\right].\nonumber 
\end{eqnarray}

Note that $A_{\mu}$ is not a dynamical field at tree level because there is no associated kinetic term in (\ref{lfim}). Nevertheless, as we are going to show, the gauge boson acquires dynamics by radiative corrections. Another point which must be stressed is that, in the limit $\xi \rightarrow \infty$, we recover the original Thirring model.

As pointed in ref. \cite{ito}, the fact that the Lagrangian (\ref{lfim}) has a gauge symmetry restricts the choice of the regularization schemes to be used, i.e., we only can employ the regularization schemes which preserve the gauge symmetry. This is the merit of the above construction in comparison with the na\"{\i}ve use of the Lagrangian (4) with a gauge fixing term, without the prior introduction of a gauge symmetry (see \cite{gomes}\cite{hp}). Nevertheless, this is not sufficient to remove the regularization ambiguity in the coefficient of the induced Chern-Simons term, when we calculate the fermion loop corrections. In this sense the causal method has been proven to be useful\cite{swpt}\cite{sch}, because it never runs into the usual difficulties associated to ultraviolet divergencies, resulting in a unambiguous value for the coefficient of the induced Chern-Simons term.

\section{Epstein and Glaser Theory}

\hspace{0.7cm}In the causal approach to quantum field theory the S-matrix is viewed as an operator-valued distribution,
written as

\begin{equation}
S(g) = 1 + \sum_{n=1}^\infty  \frac{1}{n!} \int  dx_1 \ldots dx_n T_n(x_1, \ldots ,x_n)g(x_1) \ldots  g(x_n).
\label{sma}
\end{equation}

\noindent
Once $T_1(x)$ is known, the S-matrix is constructed inductively, order by order, from the causal
structure. The c-number test functions in (\ref{sma}) are supposed to belong to the Schwartz space of functions of rapid decreasing, $g(x) \in {\cal S}({\rm {\bf R^3}})$. The adiabatic limit $g \rightarrow 1$ must be considered at the end of calculations.

Analogously,  the inverse S-matrix has the form

\begin{equation}
S(g)^{-1} = 1 + \sum_{n=1}^\infty  \frac{1}{n!} \int  dx_1 \ldots  dx_n {\tilde{T}}_n(x_1, \ldots ,x_n)g(x_1) \ldots  g(x_n) ,
\end{equation}

\noindent
where ${\tilde{T}}_n(x)$ can be obtained  by formal inversion of (\ref{sma}). Since the $n$-point funtion $T_n$ (${\tilde{T}}_n $) is symmetrical in its arguments we will use the notation $X=\{ x_1,...,x_2\} $.

The inductive step is as follows: if  all  $T_m(X)$, $m \leq n-1$ are known, one can define the distributions

\begin{eqnarray}
A_n^{'}(x_1, \ldots ,x_n)&=& \sum_{P_2} {\tilde{T}}_{n_1}(X) T_{n-n_1}(Y,x_n), \nonumber \\  \\
R_n^{'}(x_1, \ldots ,x_n)&=& \sum_{P_2} T_{n-n_1}(Y,x_n) {\tilde{T}}_{n_1}(X), \nonumber
\label{arli}
\end{eqnarray}

\noindent
where the sums run over all partitions

\begin{equation}
P_2 : \{ x_1, \ldots ,x_{n-1} \} = X \cup Y \; , \hspace{0.3cm} X \neq  \O  ,
\end{equation}

\noindent
into disjoint subsets with $\mid\! X\!\mid =n_1$, $\mid\! Y\! \mid \leq n-2$. If the sums are extended in order to include the empty set $X=\O $ we get

\begin{eqnarray}
A_n (x_1, \ldots ,x_n)&=& \sum_{P_2^0} {\tilde{T}}_{n_1}(X) T_{n-n_1}(Y,x_n)  \nonumber \\  \nonumber  \\
&=& A_n^{'}(x_1, \ldots ,x_n) + T_n (x_1, \ldots ,x_n), \nonumber \\ \label{ar} \\
R_n (x_1, \ldots ,x_n)&=& \sum_{P_2^0} T_{n-n_1}(Y,x_n) {\tilde{T}}_{n_1}(X)  \nonumber \\  \nonumber \\
&=& R_n^{'}(x_1, \ldots ,x_n) + T_n (x_1, \ldots ,x_n), \nonumber
\end{eqnarray}

\noindent
where $P_2^0$ stands for all partitions

\begin{equation}
P_2^0 : \{ x_1, \ldots ,x_{n-1} \} = X \cup Y \; .
\end{equation}

\noindent
We can see that $A_n$ and $R_n$, in eq. (\ref{ar}), are not known from the hypothesis of induction because contain the unknow $T_n$. Just only the difference 

\begin{equation}
D_n (x_1, \ldots ,x_n) =R^{'}_n - A^{'}_n = R_n -A_n,
\label{d}
\end{equation}

\noindent
is known.

Making use of causality it turns out that  $R_n$ has  retarded support and  $A_n$ has advanced support, i.e. 

\begin{equation}
{\rm supp} R_n (X) \subseteq \Gamma^{+}_{n-1}(x_n), \hspace{0.5cm} {\rm supp} A_n (X) \subseteq \Gamma^{-}_{n-1}(x_n),
\label{arsup}
\end{equation}

\noindent
with

\begin{eqnarray}
\Gamma^{\pm}_{n-1}(x) \equiv  \{  (x_1, \ldots ,x_{n-1}) \mid x_j \in \overline{V}^{\pm} (x),  \forall j= 1, \ldots ,n-1 \},  \nonumber \\  \\
\overline{V}^{\pm} (x) = \{  y  \mid  (y-x)^2 \geq 0 , \pm (y^0 - x^0 ) \geq 0  \}. \nonumber 
\end{eqnarray}

The\hfill distribution\hfill $D_n$\hfill has\hfill causal\hfill support,\hfill $ {\rm supp} D_n  \subseteq \Gamma^{+}_{n-1} \cup \Gamma^{-}_{n-1}$; \hfill \\ decomposing $D_n$ in advanced and retarded distributions we obtain the $T_n$ distri{\-}bution by (\ref{ar}). 

The operator-valued distributions which we shall have to split are of the form

\begin{equation}
D_n(x_1,...,x_n) = \sum_k : \prod_j \overline{\psi}(x_j) d^k_n (x_1, \ldots ,x_n) \prod_l \psi(x_l) \prod_m A(x_m) :,
\end{equation}

\noindent
where $\psi$, $\overline{\psi}$ are the free fermion fields and $A$ the free gauge boson fields. In this expression $d_n^k$ are numerical tempered distributions, $d_n^k\in  {\cal S}^{'}({\rm {\bf R}^{3n}})$, with causal support. Because of the translation invariance,  it is sufficient  to put $x_n=0$ and consider

\begin{equation}
d(x) \equiv d_n^k (x_1,\ldots ,x_{n-1},0) \in {\cal S}^{'}({\rm {\bf R^m}}),    m=3n-3.
\end{equation}

The nontrivial step is the splitting of the numerical causal distribution $d$ in the advanced and retarded  distributions $a$ and $r$, respectively. From the fact that $\Gamma^+  (0)\cap \Gamma^- (0) = \{ 0\}$ we can see that the behaviour of $d(x)$ in $x\!= 0$ is crucial in the splitting problem. For this reason, it is necessary to classify the singular distributions. With this aim we introduce the followings definitions\cite{sch}\cite{russo}:

{\bf{Definition 1.}}  The distribution $d(x) \in {\cal S}^{'} ({\bf  R}^m)$ has a quasi-asymptotics  $d_0 (x)$ at $x=0$ with respect to a positive continuous  function $\rho (\delta)$, $\delta  >0$, if the limit

\begin{equation}
\lim_{\delta \rightarrow 0} \rho (\delta) \delta^m d(\delta x)= d_0(x) \equiv\!\!\!\!\!\slash \; 0,
\label{d1}
\end{equation}

\noindent
exists in ${\cal S}^{'}({\bf R}^m)$.

\noindent
The equivalent definition in momentum space reads

{\bf{Definition 2.}}  The distribution $\hat{d}(p) \in {\cal S}^{'} ({\bf  R}^m)$ has a quasi-asymptotics  $\hat{d}_0 (p)$ at $p=\infty $ if the limit

\begin{equation}
\lim_{\delta \rightarrow 0} \rho (\delta) \langle \hat{d}(\frac{p}{\delta }), \stackrel{\vee}{\phi}(p)\rangle = \langle \hat{d}_0, \stackrel{\vee}{\phi} \rangle,
\label{d2}
\end{equation}

\noindent
exists for all $\stackrel{\vee}{\phi}\;\;\in {\cal S}({\bf R}^{\rm m})$.

In (\ref{d2})  $\hat{d} $ denotes the distributional Fourier transform of $d$ and $\stackrel{\vee}{\phi}$ the inverse Fourier transform of  $\phi $. The function $\rho (\delta )$ is called the power-counting function.

{\bf{Definition 3.}}  The distribution $d \in {\cal S}^{'} ({\bf  R}^{\rm m})$ is called singular of order $\omega$ if it has a quasi-asymptotics  $d_0 (x)$ at $x=0 $, or its Fourier transform has a quasi-asymptotics  $\hat{d}_0 (p)$ at $p=\infty $, respectively, with power-counting function $\rho (\delta )$ satisfying

\begin{equation}
\lim_{\delta \rightarrow 0} \frac{\rho ( c \delta )}{ \rho ( \delta )}= c^{\omega}, 
\label{d3}
\end{equation}

\noindent
 for  each $c > 0$. 

It follows that\cite{eg}\cite{sch} for $\omega < 0$ the solution is unique and can be defined by multiplication by a step function. For $\omega \geq 0$ the retarded distribution obtained from the causal splitting can be written down  by means of the ``dispersion'' formula\cite{sch}

\begin{equation}
\hat{r} (p) = \frac{i}{2\pi}\int_{- \infty}^{+ \infty} dt \frac{\hat{d}(tp)}{(t-i0)^{\omega +1}(1-t+i0)}.
\label{cs}
\end{equation}

\noindent
But, in contrast with the case ${\omega}<0$, this solution is not unique. If $\tilde{r} (x)$ is the retarded part of another decomposition, then $\tilde{r}(x) - r(x)$ is a distribution with support in $\{ 0 \}$. In momentum space this gives the general solution of the splitting problem

\begin{equation}
\tilde{r}(p) = \hat{r}(p) + \sum_{\mid a \mid = 0}^{\omega} C_a p^{a} ,
\label{gen}
\end{equation}

\noindent
where the constant coefficients $C_a$ are not fixed by the causal structure; additional physical conditions are needed to determine them.

Some comments about the solution (\ref{cs}) and (\ref{gen}) are noteworthy. First, the solution (\ref{cs}), called central splitting solution, preserves the symmetries of  the theory, in special Lorentz covariance and gauge invariance. Second, in expression  (\ref{gen})  it was assumed the {\it minimal distribution splitting} condition which says that  the singular order  cannot be raised in the splitting. This condition, very important in QED$_4$\cite{sch} and QED$_3$\cite{swpt}\cite{tom}, will also be useful here.  Finally, the right singular order $\omega$ in (\ref{cs}) is essential, since if we underestimate $\omega$, the integral in (\ref{cs}) will not be convergent and, again, one runs into the ultraviolet divergencies of the usual perturbation theory.

\section{Nonrenormalizability Proof}

\hspace{0.7cm}The Lagrangian ${\cal L}_{A,\psi}$, eq. (\ref{lfim}), is our starting point for the causal treatment of the Thirring model as a gauge theory. From (\ref{lfim}) we see that the first order term in the causal perturbative expansion of the S-matrix is

\begin{equation}
T_1(x)=-ie:\overline{\psi}(x)\gamma^{\mu}\psi(x):A_{\mu}=-\tilde{T}_1(x)
\label{acopl}
\end{equation}

\noindent
From this expression we see that the dimensionless parameter $e$ plays the role of an expansion parameter, analogous to the electric charge in QED. But in the limit $\xi \rightarrow \infty$, when we recover the relation between $\tilde{A}_{\mu}$ and the fermion current in the original Thirring model, the true expansion parameter is $G$.

As pointed out in the last section, the $A_{\mu}$ field is not a dynamical field (the genuine dynamical field is $\partial_{\mu}A^{\mu}$). Therefore, there is no propagation associated to $A_{\mu}$ . Nevertheless, we can, formally, associate to $A_{\mu}$ a ``propagator". From (\ref{lfim}) we obtain the Feynman Green's function

\begin{equation}
D_{\mu\nu}^F(k)=\frac{i}{\sqrt{2\pi}}\frac{1}{M^2}\left( g_{\mu\nu}-\frac{k_{\mu}k_{\nu}}{k^2-\xi M^2}\right).
\label{propag}
\end{equation}

\noindent
In the same way we get the commutation functions

\begin{equation}
D_{\mu\nu}^{(\pm )}(x)=\pm \frac{i}{(2\pi)^2}\int d^3k \frac{k_{\mu}k_{\nu}}{M^2}\delta (k^2-\xi M^2)\theta (\pm k_0)e^{-ik\cdot x}.
\label{fcom}
\end{equation}

\noindent
This enable us to obtain the second order distributions $T_2$. However, before starting the perturbation theory, it is useful to derive a general expression for the singular order $\omega$ of arbitrary graphs.

{\bf Proposition 1}. For the Thirring model the singular order is

\begin{equation}
\omega=3-f-\frac{3}{2}b+\frac{1}{2}n,
\label{ordem}
\end{equation}

\noindent
where $f$ ($b$) is the number of external fermions (bosons) and $n$ is the order of perturbation theory.

{\it Proof}. The proof is by induction\cite{sch}\cite{swpt}. First we verify (\ref{ordem}) for the diagrams in lowest order. The first order term (\ref{acopl}) has $\omega=0$, by definition. 

Then, to verify that this relation is preserved in the step from $n-1$ to $n$ in perturbation theory, we must consider a tensor product of  two subgraphs with singular order $\omega_1$ and $\omega_2$ which satisfies (\ref{ordem}), by hypothesis. This tensor product has to be normally ordered, giving rise to bosonic and fermionic contractions. Here we will consider only the bosonic case, since fermionic contractions have already been considered in the context of  QED$_3$\cite{swpt}\cite{tom}.
 
Suppose that $l$ bosonic contractions arise in the process. Then the numerical distribution of the contracted expression is

\begin{eqnarray}
t_1^{[\mu]}(x_1\!-\!x_r,\ldots,x_{r-1}\!-\!x_r)\prod_{j=1}^{l}D_{\mu_j\nu_j}^{(+)}(x_{r_j}\!-\!y_{v_j})t_2^{[\nu]}(y_1\!-\!y_v,\ldots,y_{v-1}\!-\!y_v) \nonumber \\ \nonumber \\
\equiv t(\zeta_1,\ldots,\zeta_{r-1},\eta_1,\ldots,\eta_{v-1},\eta),
\label{defcon}
\end{eqnarray}

\noindent
where we have taken into account the translation invariance. In this expression $\{x_{r_j}\}$ is a subset of $\{x_1,\ldots,x_r\}$ and $\{y_{v_j}\}$ is a subset of $\{y_1,\ldots,y_v\}$. We have introduced the relative coordinates

\begin{equation}
\zeta_j=x_j-x_r,\hspace{0.4cm}\eta_j=y_j-y_v,\hspace{0.4cm}\eta=x_r-y_v,
\end{equation}

\noindent
and the superscripts $[\mu]$ and $[\nu]$ means the collection of  indices $\{\mu_1,\ldots,\mu_l\}$ and $\{\nu_1,\ldots,\nu_l\}$, respectively. 

The Fourier transform of $t({\bf \zeta,\eta})$ in (\ref{defcon}), taking into account that products go into convolutions, is

\begin{eqnarray}
\hat{t}(p_1,\ldots,p_{r-1},q_1,\ldots,q_{v-1},q)\propto\int \left(\prod_{j=1}^{l}d^3k_j\right)\delta^{(3)}(q-\sum_{j=1}^l k_j) \nonumber \\ \nonumber \\
\times \hat{t}_1^{[\mu]}(\ldots,p_i-k_{r(i)},\ldots)\prod_{j=1}^l \hat{D}_{\mu\nu}^{(+)}(k_j) \hat{t}_2^{[\nu]}(\ldots,q_s+k_{v(s)},\ldots),
\end{eqnarray}

\noindent
where $r(i)=v(s)$ if and only if $x_i$ and $y_s$ are joined by a contraction. For the coordinates $x_j$ and $y_m$ which are not joined by a contraction we have just $p_j$ and $q_m$ as arguments, respectively. The proportionality sign is to indicate that we are omitting powers of $2\pi$.

Applying $\hat{t}(p_1,\ldots,q)$ in a test function $\stackrel{\vee}{\phi}\; \in{\cal S}({\rm R}^{3(r+v-1)})$ we get, after some algebra,

\begin{equation}
\langle\hat{t},\stackrel{\vee}{\phi}\rangle\propto\int d^{3r-1}p^{'} d^{3v}q^{'}\hat{t}_1^{[\mu]}(p^{'})\hat{t}_2^{[\nu]}(q^{'})\psi_{[\mu\nu]}(p^{'},q^{'}),
\end{equation}

\noindent
where $\psi_{[\mu\nu]}(p^{'},q^{'})$ is defined as

\begin{eqnarray}
\psi_{[\mu\nu]}(p^{'},q^{'})&=&\int\left(\prod_{j=1}^l d^3k_j\right)d^3q^{'}\delta^{(3)}(q^{'}-\sum_{j=1}^l k_j) \prod_{j=1}^l \hat{D}_{\mu\nu}^{(+)}(k_j)  \nonumber \\ \nonumber \\
&\times& \stackrel{\vee}{\phi}(\ldots,p_i^{'}+k_{r(i)},\ldots,q^{'}_s-k_{v(s)},\ldots,q^{'}).\label{defpsimunu}
\end{eqnarray}

In order to determine the singular order of $\hat{t}$, according to definition 2, we have to consider the scaled distribution $\hat{t}(\frac{p_1}{\delta},\ldots, \frac{q}{\delta})$. Then, we find

\begin{equation}
\langle\hat{t}(\frac{p_1}{\delta},\ldots, \frac{q}{\delta}),\stackrel{\vee}{\phi}\rangle=\delta^m \int d^{3r-3}p^{'} d^{3v-3}q^{'} \hat{t}_1^{[\mu]}(p^{'})\hat{t}_2^{[\nu]}(q^{'})\psi_{\delta[\mu\nu]}(p^{'},q^{'}),
\label{ditesc}
\end{equation}

\noindent
with $m=3(r+v-1)$ and

\begin{eqnarray}
\psi_{\delta[\mu\nu]}(p^{'},q^{'})&=&\int\left(\prod_{j=1}^l d^3k_j\right)d^3q^{'}\delta^{(3)}(q^{'}-\sum_{j=1}^l k_j) \prod_{j=1}^l \hat{D}_{\mu\nu}^{(+)}(k_j) \nonumber \\ \nonumber \\
&\times& \stackrel{\vee}{\phi}(\ldots,\delta(p_i^{'}+k_{r(i)}),\ldots,\delta(q^{'}_s-k_{v(s)}),\ldots,\delta q^{'}).
\end{eqnarray}

\noindent
We intoduce the new variables $\tilde{k}_j=\delta k_j$ and $\tilde{q}=\delta q$, and observing that

\begin{eqnarray}
\hat{D}^{(+)}_{\mu\nu}(\frac{\tilde{k}}{\delta})&=&\frac{\tilde{k}_{\mu}\tilde{k}_{\nu}}{\delta^2 M^2} \delta^{(1)}(\frac{\tilde{k}^2}{\delta^2}-\xi M^2)\theta(\frac{\tilde{k}_0}{\delta})\nonumber \\ \nonumber \\
&=&\frac{\tilde{k}_{\mu}\tilde{k}_{\nu}}{M^2}\delta^{(1)}(\tilde{k}^2)\theta(k_0) \equiv \hat{D}^{(+)}_{0 \mu\nu}(\tilde{k}),
\end{eqnarray}

\noindent
we obtain

\begin{equation}
\psi_{\delta [\mu\nu]}(p,q)=\frac{1}{\delta^{3l}}\psi_{[\mu\nu]}^0({\delta}p,{\delta}q),
\end{equation}

\noindent
where the superscript $0$ indicates that $\hat{D}^{(+)}_{\mu\nu}$ is replaced by $\hat{D}^{(+)}_{0 \mu\nu}$ in (\ref{defpsimunu}). Then, using this result and $\delta p=\tilde{p}$, $\delta{q}=\tilde{q}$, we find from (\ref{ditesc})

\begin{equation}
\langle\hat{t}(\frac{p_1}{\delta},\ldots ,\frac{q}{\delta}),\stackrel{\vee}{\phi}\rangle=\delta^{3-3l} \int d^{3r-3}\tilde{p} d^{3v-3}\tilde{q} \hat{t}_1^{[\mu]}(\frac{\tilde{p}}{\delta})\hat{t}_2^{[\nu]}( \frac{\tilde{q}}{\delta}) \psi_{[\mu\nu]}^0(\tilde{p},\tilde{q}).
\end{equation}

\noindent
But, by the induction hypothesis, the distributions $\hat{t}_1^{[\mu]}$ and $\hat{t}_2^{[\nu]}$ have singular orders $\omega_1$ and $\omega_2$ with power-counting functions  $\rho_1(\delta)$ and $\rho_2(\delta)$, respectively. So we verify that the limit considered in definition 2 exists for the distribution $\hat{t}$ with power-counting function given by

\begin{equation}
\rho(\delta)=\delta^{3l-3}\rho_1(\delta)\rho_2(\delta),
\end{equation}

\noindent
with singular order

\begin{equation}
\omega =3l-3+\omega_1+\omega_2.
\end{equation}

\noindent
Thus, substituting

\begin{equation}
\omega_i=3-f_i-\frac{3}{2}b_i+\frac{1}{2}n_i,
\end{equation}

\noindent
for $\omega_1$ and $\omega_2$ gives

\begin{equation}
\omega=3-(f_1+f_2)-\frac{3}{2}(b_1+b_2-2l)+\frac{1}{2}(n_1+n_2),
\end{equation}

\noindent
that proves the above proposition for $l$ bosonic contractions.

From eq. (\ref{ordem}) we have that, for one loop corrections, the vacuum polarization ($n=2$, $f=0$, $b=2$) has $\omega_{vp}=1$, the fermion self-energy ($n=2$, $f=2$, $b=0$) has $\omega_{se}=2$ and the vertex correction ($n=3$, $f=2$, $b=1$), $\omega_v=1$. We consider here only the vacuum polarization tensor. The fermion self-energy and the vertex correction will be considered elsewhere\cite{nos}.

In addition, from proposition 1 it follows that the Thirring model is a nonrenormalizable theory, as expected. This means that the number of  free parameters, i.e., the coefficients of the polynomial in $p$ in eq. (\ref{gen}),  increases indefinitely when we consider higher orders in perturbation theory such that we can't fix all of  them by symmetry considerations. However, as we shall see in the next section, for the vacuum polarization tensor we will be able to determine all constants appearing in second order perturbation theory.  

\section{Dynamical Mass Generation}

\hspace{0.7cm}In this section we consider the vacuum polarization and  address the dynamical generation of a kinetic term for the gauge boson. Since the vacuum polarization tensor assumes the same form that in QED$_3$, we omit the details of the calculation, which can be found in refs. \cite{swpt}\cite{sch}.

In\hfill second-order\hfill perturbation\hfill theory\hfill we\hfill can\hfill construct\hfill the\hfill distribution \\ $D_2(x_1,x_2)=R_2^{'}-A_2^{'}$ following the steps outlined in the section 1. So, using Wick's theorem, the contribution for the vacuum polarization in $D_2$ is

\begin{eqnarray}
D_{2 vp}(x_1,x_2)&=&-e^2 tr\left[\gamma^{\mu}S^{(-)}(y)\gamma^{\nu}S^{(+)}(-y)- \gamma^{\mu}S^{(+)}(y)\gamma^{\nu}S^{(-)}(-y)\right] \nonumber \\ \nonumber \\
&\times&:A_{\mu}(x_1)A_{\nu}(x_2):,
\label{vacuo}
\end{eqnarray}

\noindent
where $y\equiv x_1-x_2$ and 

\begin{equation}
S^{(\pm )}(x)=\pm\frac{i}{(2\pi)^2}\int d^3p(p\!\!\!\slash+m)\theta(\pm p_0)\delta (p^2-m^2)e^{-ip\cdot x}.
\label{comfer}
\end{equation}

The numerical distribution associated with $D_2(x_1,x_2)$ can be written in the form

\begin{equation}
d^{\mu\nu}(x_1,x_2)=P^{\mu\nu}(y)-P^{\nu\mu}(-y),
\label{dmunu}
\end{equation}

\noindent
where

\begin{equation}
P^{\mu\nu}(y)\equiv e^2 {\rm tr}\left[\gamma^{\mu}S^{(+)}(y)\gamma^{\nu}S^{(-)}(-y)\right].
\label{pmunu}
\end{equation}

The distribution $d^{\mu\nu}(y)$ can be shown to have causal support\cite{swpt}\cite{sch}. So we can proceed  the  splitting  according to the procedure already explained, following closely ref. \cite{swpt}. In momentum space, using (\ref{comfer}) for the fermion commutation functions, $P^{\mu\nu}(k)$ can be written as

\begin{equation}
\hat{P}^{\mu\nu}(k)=-\frac{e^2}{(2\pi)^{\frac{5}{2}}}\int d^3p \; \theta (p_0)\delta (p^2-m^2)\theta (k_0-p_0) \delta [(k-p)^2-m^2]j^{\mu\nu}(k,p),
\label{distpmunu}
\end{equation}
 
\noindent
with 

\begin{eqnarray}
j^{\mu\nu}(k,p)&=&{\rm tr}\left[\gamma^{\mu}(p\!\!\!\slash+m)\gamma^{\nu}(k\!\!\!\slash-p\!\!\!\slash-m)\right] \nonumber \\ \nonumber \\
&=&-2[(m^2-p^2)g^{\mu\nu}+2p^{\mu}p^{\nu}-( p^{\mu}k^{\nu}+ k^{\mu}p^{\nu})\nonumber \\
&+&g^{\mu\nu}p\cdot k +im\varepsilon^{\mu\nu\delta}k_{\delta}].
\label{jota}
\end{eqnarray}

\noindent
From (\ref{distpmunu}) and (\ref{jota}) one can observe that $P^{\mu\nu}$ is gauge invariant

\begin{equation}
k_{\mu}\hat{P}^{\mu\nu}(k)=0.
\end{equation}

\noindent
This property enable us to attribute to $P^{\mu\nu}$ the following tensor structure

\begin{equation}
\hat{P}^{\mu\nu}(k)= \hat{P}^{\mu\nu}_s(k)+ \hat{P}^{\mu\nu}_a(k),
\end{equation}

\noindent
with

\begin{eqnarray}
\hat{P}^{\mu\nu}_s(k)= (k^{\mu}k^{\nu}-k^2g^{\mu\nu})\tilde{B}_1(k^2),   \\ \nonumber \\
\hat{P}^{\mu\nu}_a(k)= im\varepsilon^{\mu\nu\delta}k_{\delta}\tilde{B}_2(k^2).
\end{eqnarray}

Projecting out $\tilde{B}_1(k^2)$ and $\tilde{B}_2(k^2)$ from $\hat{P}^{\mu\nu}(k)$, one obtains

\begin{equation}
\tilde{B}_1(k^2)=-\frac{e^2}{(2\pi)^{\frac{3}{2}}}\frac{1}{8\sqrt{k^2}}\left(1+\frac{4m^2}{k^2}\right)\theta (k_0)\theta (k^2-4m^2),
\label{tb1}
\end{equation}

\noindent
and

\begin{equation}
\tilde{B}_2(k^2)= \frac{e^2}{(2\pi)^{\frac{3}{2}}}\frac{1}{2\sqrt{k^2}}\theta (k_0)\theta (k^2-4m^2).
\label{tb2}
\end{equation}

Since the Fourier transform of $P^{\mu\nu}(-y)$ is given by $\hat{P}^{\mu\nu}(-k)$, from (\ref{dmunu}) we see that

\begin{eqnarray}
\hat{d}^{\mu\nu}(k)&=&\hat{P}^{\mu\nu}(k) - \hat{P}^{\nu\mu}(-k) \nonumber \\ \nonumber \\
&=& \hat{d}^{\mu\nu}_s(k)+ \hat{d}^{\mu\nu}_a(k),
\end{eqnarray}

\noindent
where

\begin{eqnarray}
\hat{d}^{\mu\nu}_s(k)&=& (k^{\mu}k^{\nu}-k^2g^{\mu\nu})B_1(k^2), \label{partesim}  \\ \nonumber \\
\hat{d}^{\mu\nu}_a(k)&=& im\varepsilon^{\mu\nu\delta}k_{\delta}B_2(k^2), \label{parteantisim}
\end{eqnarray}

\noindent
with $ B_1(k^2)$ and $ B_2(k^2)$ given by expressions (\ref{tb1}) and (\ref{tb2}), replacing $\theta(k_0)$ by ${\rm sgn}(k_0)$. 

From eqs. (\ref{partesim}) and (\ref{parteantisim}) we find that the singular order of $\hat{d}^{\mu\nu}_s(k)$ and $\hat{d}^{\mu\nu}_a(k)$ are $\omega_s=1$ and $\omega_a=0$, respectively. Then the distribution splitting is non-trivial ($\omega \geq 0$) and we need to use the central splitting solution, eq. (\ref{cs}), with the appropriate $\omega$ to obtain the retardaded distribution. Since $\hat{d}^{\mu\nu}_s$ and $\hat{d}^{\mu\nu}_a$ are independent due to the tensor structure, the splitting process for each one must be considered in separate.

For the symmetric part  we have 

\begin{equation}
\hat{r}^{\mu\nu}_s(k) = \frac{i}{2\pi}(k^{\mu}k^{\nu}-k^2g^{\mu\nu})\int_{- \infty}^{+ \infty} dt \frac{t^2B_1(t^2k^2)}{(t-i0)^{2}(1-t+i0)},
\end{equation}

\noindent
which results

\begin{equation}
\hat{r}^{\mu\nu}_s(k)= \frac{i}{(2\pi)^{\frac{3}{2}}}\left( g^{\mu\nu}-\frac{k^{\mu}k^{\nu}}{k^2}\right) \Pi^{(1)}(k^2),
\label{simetr}
\end{equation}

\noindent
with

\begin{eqnarray}
\Pi^{(1)}(k^2)
&=& \frac{e^2}{16\pi}k^2{\rm sgn}(k_0) \left[\frac{4m^2}{k^2}+\frac{1}{\sqrt{k^2}}\left(1+\frac{4m^2}{k^2}\right) \right. \nonumber \\ \nonumber \\
&\times& \left. \left( \ln \left| \frac{1-\sqrt{\frac{k^2}{4m^2}}}{1+\sqrt{\frac{k^2}{4m^2}}}\right|- i\pi\theta (k^2-4m^2)\right) \right]. 
\label{pi1}
\end{eqnarray}

Since the singular order of $\hat{d}^{\mu\nu}_s$ is $\omega_s=1$, from (\ref{gen}) we have for the general solution of the splitting

\begin{equation}
\tilde{r}^{\mu\nu}_s(k)=\hat{r}^{\mu\nu}_s(k) +C_0 g^{\mu\nu}+C_{\delta}'\varepsilon^{\mu\nu\delta}+C_1^{\mu}k^{\nu}+C_2^{\nu}k^{\mu},
\label{rsgen}
\end{equation}

\noindent
where $C_0$, $C_{\delta}'$, $C_1^{\mu}$ and $C_2^{\nu}$ are constants which are not fixed by causality. However, $C_{\delta}'=0$ to preserve the symmetric structure. The Lorentz structure and the fact that $C_1^{\mu}$ and $C_2^{\nu}$ are c-numbers lead to $C_1^{\mu}=C_2^{\nu}=0$. By the requirement of gauge invariance, $k_{\mu}\tilde{r}^{\mu\nu}_s(k)=0$, $C_0$ must vanish. So, the general solution of the splitting problem for the symmetrical part is given by (\ref{simetr}).

For the antisymmetric part we use eq. (\ref{cs}) with $\omega=0$

\begin{equation}
\hat{r}^{\mu\nu}_a(k) = -\frac{m}{2\pi}\varepsilon^{\mu\nu\delta}k_{\delta}\int_{- \infty}^{+ \infty} dt \frac{tB_2(t^2k^2)}{(t-i0)(1-t+i0)},
\end{equation}

\noindent
from which we obtain

\begin{equation}
\hat{r}^{\mu\nu}_a(k) = -\frac{m}{(2\pi)^{\frac{3}{2}}} \varepsilon^{\mu\nu\delta}k_{\delta}\Pi^{(2)}(k^2),
\label{antiss}
\end{equation}

\noindent
where

\begin{equation}
\Pi^{(2)}(k^2) = \frac{e^2}{4\pi}\frac{{\rm sgn} (k_0)}{\sqrt{k^2}}\left[ \ln \left| \frac{1-\sqrt{\frac{k^2}{4m^2}}}{1+\sqrt{\frac{k^2}{4m^2}}}\right|- i\pi\theta (k^2-4m^2) \right].
\label{pi2}
\end{equation}

The general solution for the antisymmetric part is

\begin{equation}
\tilde{r}^{\mu\nu}_a(k)={\hat{r}}^{\mu\nu}_a(k) +C_0 g^{\mu\nu}+C_{1\delta}\varepsilon^{\mu\nu\delta}.
\label{genant}
\end{equation}

\noindent
However, the constant $C_0$ must vanish to preserve the antisymmetric structure and $C_{1\delta}$ must also vanish by gauge invariance, so that the general solution of the splitting of antisymmetric part is given by (\ref{antiss}).

In this point, is interesting to note that, in spite of the fact that the model is nonrenormalizable, we were able to determine all constants $C_a$ appearing in the general solution for the polarization tensor. Of course, this is not always the case. In fact, for the fermion self-energy and the vertex correction there remains one undetermined constant\cite{nos}.

The vacuum polarization tensor is defined as

\begin{equation}
\Pi_{\mu\nu}(k)=-i(2\pi)^{\frac{3}{2}} \left( {\hat{r}}_{\mu\nu}(k)- {\hat{r}'}_{\mu\nu}(k) \right),
\end{equation}

\noindent
where ${\hat{r}'}_{\mu\nu}(k)$ is the Fourier transform of the numerical distribution associated with $R_2^{'}(x_1,x_2)$, eq. (\ref{arli}). In this case, ${\hat{r}'}_{\mu\nu}(k)=-\hat{P}_{\mu\nu}(-k)$ and, from (\ref{pmunu}), we see that this distribution do not contribute in the region $k^2 < 4m^2$. Thus we can write down the polarization tensor as

\begin{equation}
\Pi^{\mu\nu}(k)=\left(g^{\mu\nu}-\frac{k^{\mu}k^{\nu}}{k^2}\right)\Pi^{(1)}(k^2)+im\varepsilon^{\mu\nu\delta}k_{\delta}\Pi^{(2)}(k^2),
\label{polageral}
\end{equation}

\noindent
with $\Pi^{(1)}(k^2)$ and $\Pi^{(2)}(k^2)$ given by eqs. (\ref{pi1}) and (\ref{pi2}), respectively, satifying

\begin{eqnarray}
\Pi^{(1)}(0)&=&0, \nonumber \\ \label{piem0} \\
\Pi^{(2)}(0)&=&-\frac{e^2}{4\pi m}. \nonumber
\end{eqnarray}

Let us now derive the gauge boson propagator modified by vacuum polarization insertions, in the one loop approximation. This is given by the series

\begin{equation}
{\cal D}= D_F+iD_F\Pi D_F +iD_F\Pi D_F\Pi D_F +\ldots =D_F+iD_F\Pi{\cal D},
\end{equation}

\noindent
from which

\begin{equation}
{\cal D}^{-1}_{\mu\nu}= (D^F_{\mu\nu})^{-1}-i\Pi_{\mu\nu},
\label{propcompleto}
\end{equation}

\noindent
where $D_F$ is the free gauge boson propagator, eq. (\ref{propag}).

Here, it is important to note that, by the Coleman-Hill theorem\cite{coh}, this approximation for the gauge boson propagator gives the exact contribution for the topological mass term. In what follows we reproduce the Coleman-Hill argument  in the context of distribution theory\cite{gel}.

Let us consider a $n$-gauge boson ``effective vertex'' given by the sum of all graphs consisting of a single closed fermion loop with $n$ external gauge bosons attached. Associated to this vertex we have a numerical regular distribution $\hat{t}_{\mu_1 \ldots \;\mu_n}(k_1,\ldots ,k_n)$, a generalized function of the $n-1$ independent momenta. By convention, we will take the first $n-1$ momenta as the independent ones and $k_n$ fixed by momentum conservation. We will consider the distribution  $\hat{t}_{\mu_1 \ldots \;\mu_n}$ in Euclidean space, where it is an analytic generalized function of the momenta. So, gauge invariance entails

\begin{equation}
k_1^{\mu_1} \hat{t}_{\mu_1 \ldots \;\mu_n}(k_1,\ldots ,k_n)= 0.
\end{equation}

\noindent
Differentiating with respect to $k_1^{\nu}$ and taking $k_1^{\nu}=0$, we get

\begin{equation}
\hat{t}_{\mu_1 \ldots \;\mu_n}(0,k_2,\ldots ,k_n) = 0,
\end{equation}

\noindent
or, expanding in Taylor series (remember that $\hat{t}_{\mu_1 \ldots \;\mu_n}$ is a regular distribution)

\begin{equation}
\hat{t}_{\mu_1 \ldots \;\mu_n}(k_1,\ldots ,k_n) = {\cal O}(k_1).
\label{ordemk}
\end{equation}

In the same way, $\hat{t}_{\mu_1 \ldots \;\mu_n}$ is also ${\cal O}(k_2)$. Since, for $n>2$, $k_1$ and $k_2$ are independent variables, we have

\begin{equation}
\hat{t}_{\mu_1 \ldots \;\mu_n}(k_1,\ldots ,k_n) = {\cal O}(k_1k_2), \hspace{1.2cm} n>2.
\label{ordemkk}
\end{equation}

This shows that $\hat{t}_{\mu_1 \ldots \;\mu_n}$ must be, at least, ${\cal O}(k_1 \ldots k_{n-1})$. However, by Lorentz structure, it turns out that $\hat{t}_{\mu_1 \ldots \;\mu_n}$ is ${\cal O}(k_1 \ldots k_n)$ for $n>2$ (see appendix of ref.  \cite{coh}).

Then, we can construct a gauge boson self-energy graph contracting bosonic lines of fermion loops (this implies contracting $\hat{t}_{\mu_1 \ldots \;\mu_n}$ with the commutation function $D_{\mu\nu}^{(+)}(k)$, eq. (\ref{fcom})). Contracting all lines of a graph but two, which are the external lines of the graph carrying momenta $k$ and $-k$, we have three possibilities: $i$) the two external lines are attached to distinct loops; $ii$) the two external lines end at the same loop, but this has more than two bosonic lines; $iii$) the two external lines end at the same loop and this has only two bosonic lines. In cases $(i)$ and $(ii)$ the corresponding distributions are ${\cal O}(k^2)$ due to (\ref{ordemk}) and (\ref{ordemkk}), respectively. But from (\ref{polageral}) and (\ref{propcompleto}) we see that the topological mass is given by the coefficient of the term linear in $k$ when $k^2\rightarrow 0$. So, the only contribution to the topological mass comes from the second order perturbation theory, case $(iii)$ above. 

One can see that gauge invariance plays a central role in the derivation of this  result and, again, we see the relevance of the construction outlined in section 1\cite{ito}\cite{kon}. In addition, we must observe that in this theory there is no infrared difficulties because the gauge bosons are massive.

Let us now turn to the inversion of (\ref{propcompleto}). This is more easily performed by introducing the following projection operators

\begin{eqnarray}
P^{\mu\nu}_{(1)}&=& \frac{1}{2}\left( g^{\mu\nu}-\frac{k^{\mu}k^{\nu}}{k^2}+i\varepsilon^{\mu\nu\delta}\frac{k_{\delta}}{\sqrt{k^2}}\right), \nonumber \\ \nonumber \\
P^{\mu\nu}_{(2)}&=& \frac{1}{2}\left( g^{\mu\nu}-\frac{k^{\mu}k^{\nu}}{k^2}-i\varepsilon^{\mu\nu\delta}\frac{k_{\delta}}{\sqrt{k^2}}\right),  \\ \nonumber \\
P^{\mu\nu}_{(3)}&=&\frac{k^{\mu}k^{\nu}}{k^2}.\nonumber
\label{projetores}
\end{eqnarray}

\noindent
This orthonormal set of operators satisfies the relation

\begin{equation}
\sum_{i=1}^3 P^{\mu\nu}_{(i)}=g^{\mu\nu}.
\end{equation}

Then, the free gauge boson propagator can be written as a linear combination of these projection operators as 

\begin{equation}
D_F^{\mu\nu}=\frac{i}{M^2}\left( P^{\mu\nu}_{(1)} + P^{\mu\nu}_{(2)} -\frac{\xi M^2}{k^2-\xi M^2} P^{\mu\nu}_{(3)} \right),
\end{equation}

\noindent
such that the inversion is a trivial task. We just write down its inverse,

\begin{equation}
(D_F^{\mu\nu})^{-1}=-iM^2\left( P^{\mu\nu}_{(1)} + P^{\mu\nu}_{(2)} -\frac{k^2-\xi M^2}{\xi M^2} P^{\mu\nu}_{(3)} \right).
\end{equation}

\noindent
In the same way we write the polarization tensor as

\begin{equation}
\Pi^{\mu\nu}(k)=( P^{\mu\nu}_{(1)}+ P^{\mu\nu}_{(2)})\Pi^{(1)}(k^2) +m\sqrt{k^2}( P^{\mu\nu}_{(1)}- P^{\mu\nu}_{(2)})\Pi^{(2)}(k^2).
\end{equation}

\noindent
Introducing these expressions in (\ref{propcompleto}) we get

\begin{eqnarray}
{\cal D}^{\mu\nu}&=&-\frac{i}{k^2-\tilde{\Pi}(k^2)}\left[ \left( g^{\mu\nu}-\frac{k^{\mu}k^{\nu}}{k^2}\right)\frac{M^2+\Pi^{(1)}(k^2)}{[m\Pi^{(2)}(k^2)]^2}-     i\varepsilon^{\mu\nu\delta}\frac{k_{\delta}}{m\Pi^{(2)}(k^2)}\right] \nonumber \\ \nonumber \\
&-&i\xi\frac{k^{\mu}k^{\nu}}{k^2(k^2-\xi M^2)},
\label{propcomp}
\end{eqnarray}

\noindent
where

\begin{equation}
\tilde{\Pi}(k^2)=\frac{(M^2+\Pi^{(1)}(k^2))^2}{[m\Pi^{(2)}(k^2)]^2}.
\end{equation}

The form of the corrected propagator indicates that a pole is generated by the fermion loop insertions, that is, the gauge boson acquires a dynamical mass through the loops effects. Calling $M_{gb}$ the mass of the gauge boson,  we see that it is given by the solution of the transcendental equation

\begin{equation}
\left( mM_{gb}\Pi^{(2)}(M_{gb}^2)\right)^2=\left( M^2+\Pi^{(1)}(M_{gb}^2)\right)^2, \hspace{0.3cm}{\rm for}\;\; 0\leq M_{gb}<2m.
\label{polo}
\end {equation}

\noindent
Before continuing the analysis of the equation above we should observe that the limit $G \rightarrow \infty$, which must be taken with $e$ fixed,  is well defined in a general gauge (although it is ill-defined in the unitary gauge, $\xi \rightarrow \infty$),  as we can see from eqs. (\ref{propag}) and (\ref{propcomp})\cite{ito}\cite{kon}. Thus we verify that there is a solution of the equation (\ref{polo}) for all $G$ for $G>0$. In special, for $G\rightarrow \infty$ we get $M_{gb}=0$, as we can see by noticing that, in this limit, $M\rightarrow 0$ and, therefore,

\begin{equation}
\tilde{\Pi}(k^2)_{M^2=0}=\frac{(\Pi^{(1)}(k^2))^2}{[m\Pi^{(2)}(k^2)]^2},
\end{equation}

\noindent
while, from (\ref{piem0}), we see that $\tilde{\Pi}(0)_{M^2=0}=0$, so that $M_{gb}=0$ is, in fact, a solution.

On the other side, for $G\rightarrow 0_+$ we have

\begin{equation}
\frac{M_{gb}}{2m}=1-\alpha \exp \left(-\frac{2\pi}{mG}\right),
\end{equation}

\noindent
where $\alpha = 2e^{-\frac{1}{2}}$.

The solution of the transcendental equation (\ref{polo}) in these two cases are consistent with the results expected. The $A_{\mu}$ field can be thought as a bound state of two fermions. So, in the limit of  very weak interaction, we expect that the mass of the vector channel is $2m$. On the other hand, in the limit of strong interaction we expect that the mass of $A_{\mu}$ goes to zero.

Finally, we should observe that the expression for the corrected propagator (\ref{propcomp}) has a well defined limit $m\rightarrow 0$, but in this case there is no pole for time-like momentum.

\section{Conclusions}

\hspace{0.7cm}In this paper we have studied the massive gauged Thirring model in the context of the Epstein and Glaser's causal theory, and derived a proof of the nonrenormalizability of the model, obtaining a general expression for the singular order of the distributions associated with an arbitrary process.

In the sequence, we have obtained the vacuum polarization tensor  by using the causal theory and have shown that the gauge boson, which at three level is an auxiliary field, becomes dynamical. The causal method naturally afforded the correct number of subtractions for the antisymmetric part of the vacuum polarization tensor and enabled us to determine the coefficient of the induced Chern-Simons term without ambiguity. Yet, the existence of a gauge symmetry leads to a result exact in $e^2\equiv GM^2$, according to Coleman-Hill theorem.

We have also solved the transcendental equation for the pole of the gauge boson in the opposite cases of very strong and very weak coupling, obtaining results in accordance with the ones of refs. \cite{kon} and \cite{hands}.

Finally, it is important to note that in applying the causal method we never run into the usual ultraviolet divergencies. Therefore, for nonrenormalizable models there is no necessity of a cut-off. However, even for renormalizable or super-renormalizable theories\cite{sch}, we still have constants which are not fixed by causality. In the case of renormalizable and super-renormalizable theories these constants are determined by physical requirements, while for those nonrenormalizable there remains a number of undetermined constants, which increases with the order of perturbation theory.

\vspace{1.5cm}
{\bf {\large Acknowledgements}}
\vspace{0.5cm}

L. A. M. is supported by Conselho Nacional de Desenvolvimento Cient\'{\i}fico e Tecnol\'{o}gico (CNPq); B. M. P. and J. L. T. are partially supported by CNPq.

\pagebreak

\end{document}